\documentclass[letterpaper, 10 pt, conference]{ieeeconf}  
\IEEEoverridecommandlockouts                              

\usepackage{graphics}    
\usepackage{times}       
\usepackage{amsmath}     
\usepackage{amssymb}     
\usepackage{graphicx}
\usepackage{tabularx}
\usepackage{booktabs}
\usepackage{multirow}
\usepackage{algorithm}
\usepackage{algorithmic}
\usepackage{float}

\usepackage[separate-uncertainty=true,number-unit-product=\ ]{siunitx}%
\usepackage{subcaption}
\usepackage{tabularx}
\usepackage[protrusion=true,expansion]{microtype}
\usepackage{caption}
\usepackage[hidelinks]{hyperref} 
\usepackage{makecell}
\usepackage{flushend}
\usepackage{subcaption}
\usepackage{tikz}

\usepackage{amssymb}
\usepackage{pifont}
\newcommand{\cmark}{\ding{51}}%
\newcommand{\xmark}{\ding{55}}%

\usepackage[capitalise]{cleveref}
\usepackage[backend=biber,style=ieee,mincitenames=1,maxnames=20,maxcitenames=1,natbib=true]{biblatex}
\def\figref#1{Fig.~\ref{#1}}

\def\eqref#1{Eq.~(\ref{#1})}

\newcommand\etal{\emph{et al. }}

\usepackage[font=small]{caption}

\title{\LARGE \bf PeRoI: A Pedestrian-Robot Interaction Dataset for Learning \\ Avoidance, Neutrality, and Attraction Behaviors in Social Navigation}

\author{Subham Agrawal \and Nico Ostermann-Myrau \and Nils Dengler \and Maren Bennewitz
\thanks{ All authors are with the Humanoid Robots Lab, University of Bonn, Germany. M. Bennewitz, S. Agrawal, and N. Dengler are additionally with the Lamarr Institute for Machine Learning and Artificial Intelligence and the Center for Robotics, Bonn, Germany. This work has partially been funded by the German Federal Ministry of Research, Technology and Space (BMFTR) under the Robotics Institute Germany (RIG), grant No. 16ME0999.
 }
}

\begin{document}
\maketitle
\thispagestyle{empty} 
\pagestyle{empty}

\begin{abstract} 
Robots are increasingly being deployed in public spaces such as shopping malls, sidewalks, and hospitals, where safe and socially aware navigation depends on anticipating how pedestrians respond to their presence. 
However, existing datasets rarely capture the full spectrum of robot-induced reactions, e.g., avoidance, neutrality, attraction, which limits progress in modeling these interactions. 
In this paper, we present the Pedestrian-Robot Interaction~(PeRoI) dataset that captures pedestrian motions categorized into attraction, neutrality, and repulsion across two outdoor sites under three controlled conditions: no robot present, with stationary robot, and with moving robot. 
This design explicitly reveals how pedestrian behavior varies across robot contexts, and we provide qualitative and quantitative comparisons to established state-of-the-art datasets. 
Building on these data, we propose the Neural Robot Social Force Model~(NeuRoSFM), an extension of the Social Force Model that integrates neural networks to augment inter-human dynamics with learned components and explicit robot-induced forces to better predict pedestrian motion in vicinity of robots. 
We evaluate NeuRoSFM by generating trajectories on multiple real-world datasets. 
The results demonstrate improved modeling of pedestrian-robot interactions, leading to better prediction accuracy, and highlight the value of our dataset and method for advancing socially aware navigation strategies in human-centered environments.
\end{abstract} 

\section{Introduction}
\label{sec:intro}

Robots are increasingly being deployed in public places, such as university campuses, shopping malls, sidewalks, and transportation hubs~\cite{niemela2019social,han2023robot,weinberg2023sharing,kyrarini2021survey}. In these socially rich environments, they must not only reach their destinations safely and efficient, but also move in ways that align with social norms and expectations. 
Social navigation has therefore become a key requirement for ensuring safe and acceptable human-robot interaction in such cases, with a large body of work in this area relying on pedestrian trajectory datasets. 
These datasets are crucial for developing data-driven models that predict trajectories and ensure socially compliant path planning. 
However, most existing datasets either exclude robots from their data or fail to capture the variety of ways in which pedestrians may respond to their presence. 
As a result, current datasets do not provide sufficient information to understand and model the nuanced interactions between pedestrians and robots.

\begin{figure}
    \centering
    \includegraphics[width=\linewidth, trim= 120 45 75 30 , clip]{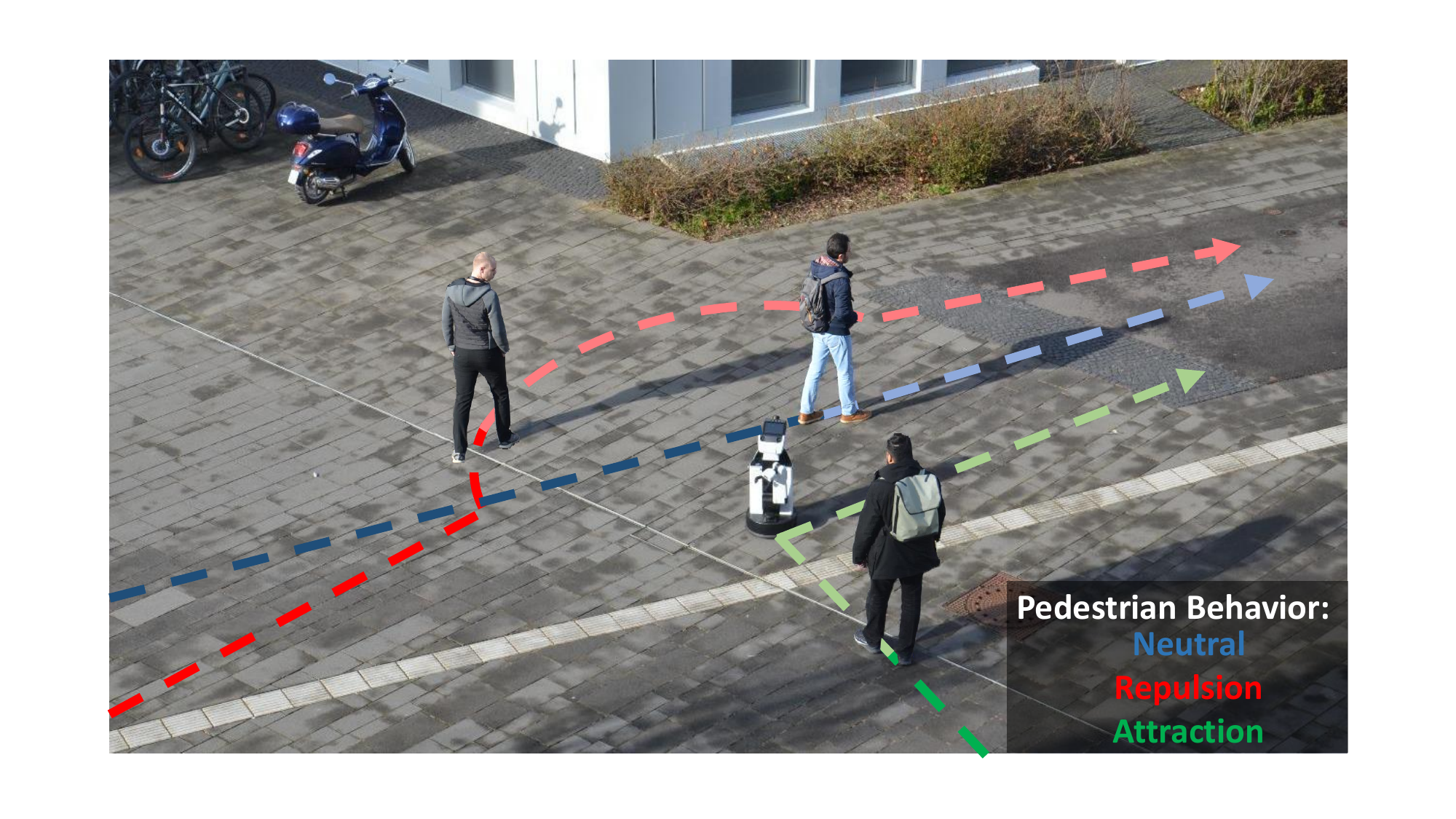}
    \caption{Example scenario of a robot influencing the trajectories of nearby pedestrians, leading them to show one of three distinct behaviors: repulsion~(red), neutral~(blue), or attraction~(green). The trajectories and the location present in the image are a visual representation of the ones in our PeRoI dataset.}
    \label{fig:overview-image}
    \vspace{-10px}
\end{figure}

This gap poses several challenges. 
Contrary to the traditional assumption that human responses to robots are uniform, real-world observations reveal that pedestrians exhibit a spectrum of behaviors~\cite{ratsamee2015social}. 
For example, some pedestrians might change their course to maintain distance from the robot~(avoidance), others might walk past the robot without any noticeable deviation in their trajectory~(neutrality), and some others might approach the robot out of curiosity~(attraction). 
Accurately capturing this diversity is critical for enabling robots to navigate naturally in public spaces. 
However, without datasets that explicitly annotate these responses, navigation policies are unable to generalize to real-world social settings.


This is also apparent from existing pedestrian trajectory prediction models. Classical approaches such as the Social Force Model~(SFM)~\cite{helbing1995social} and its extensions, as well as modern machine learning based models such as Social LSTM~\cite{alahi2016social}, DDL~\cite{wang2024pedestrian}, Trajectron++~\cite{salzmann2020trajectron++}, etc., typically either ignore robots altogether or assume that pedestrians respond to robots in a uniform way by mostly avoiding them.
These models capture interactions well in purely human-human settings but fail to differentiate between qualitatively distinct types of robot influence.
As a result, the effect of the robot is oversimplified, thus reducing the reliability of such models when deployed in real-world environments. 

To address these limitations, we introduce the \textbf{Pe}destrian-\textbf{Ro}bot \textbf{I}nteraction (PeRoI) dataset, a large-scale collection of real-world pedestrian trajectories recorded in two outdoor environments under three distinct conditions: no robot, stationary robot, and moving robot. 
Each pedestrian trajectory in the presence of a robot is annotated with one of three distinct pedestrian’s responses - avoidance, neutrality, or attraction - thereby allowing for direct study and modeling of robot-induced pedestrian dynamics, as illustrated in Fig.~\ref{fig:overview-image}.

Beyond the main dataset contribution—PeRoI—we demonstrate its utility by proposing the \textbf{Neu}ral \textbf{Ro}bot \textbf{S}ocial \textbf{F}orce \textbf{M}odel (NeuRoSFM), an extension of the classical Social Force Model (SFM) that integrates robot-induced forces and group interactions. 
By learning each force component from data using a neural network, NeuRoSFM can capture the effect of robot presence and predict diverse pedestrian behaviors more accurately. 


Our experiments show that our novel PeRoI dataset not only enhances the performance of state-of-the-art trajectory prediction models but also adds to the diversity of available pedestrian-robot interaction data. Additionally, NeuRoSFM consistently outperforms classical and optimized extensions of SFM, confirming the benefit of explicitly modeling pedestrian response to robots. Together, these contributions provide both a resource and a modeling framework that aims to advance socially aware navigation in human-centred environments.

\section{Related Work}
\label{sec:related}

\subsection{Datasets for Pedestrian-Robot Interaction}

Existing pedestrian datasets, such as ETH~\cite{pellegrini2009you}, UCY~\cite{lerner2007crowds}, and Stanford Drone~\cite{robicquet2016learning}, focus exclusively on human-human interactions. In contrast, robot-centric datasets like JRDB~\cite{martin2021jrdb}, SCAN-D~\cite{karnan2022socially}, PARSNiP~\cite{zhou2025parsnip}, SIT~\cite{bae2023sit}, and TBD~\cite{wang2024tbd} capture human motion from a robot’s perspective, but they still lack explicit labels and annotations describing pedestrian behavioral responses to the robot. As a result, although these datasets are useful for detection and tracking, they are less effective for learning socially-aware navigation models.
Some studies have explored and tried to quantify the influence of robots on pedestrian motion, for example, by measuring deviations from optimal paths~\cite{hirose2023sacson} or incorporating simple robot-repulsion terms in the SFM~\cite{agrawal2024evaluating}. Yet, these efforts focus solely on avoidance behavior while neglecting others.

In contrast, our PeRoI dataset explicitly captures and labels avoidance, neutrality, and attraction across different robot states and types. As such, our dataset enables the development of models, such as the proposed NeuRoSFM, that can predict and simulate diverse pedestrian responses, providing a more realistic base for socially-aware robot navigation research.

\subsection{Social Force Model and Extensions}

The Social Force Model (SFM)~\cite{helbing1995social} remains one of the most widely used approaches for predicting pedestrian motion, representing movement as a cumulative result of attractive and repulsive forces on a pedestrian. 
Traditionally, these include attraction toward a goal and repulsion from other pedestrians as well as obstacles. 
While computationally efficient and interpretable, the SFM in its original form assumes that interactions with other agents, such as robots, are purely repulsive and does not account for neutral or attractive behaviors. 
Beyond SFM, other classical rule-based and mathematical models have also been used for trajectory prediction, such as the Velocity Obstacle approach~\cite{large2002using}, which implements collision-avoidance constraints in velocity space, and Continuum theory~\cite{hughes2002continuum}, which models pedestrian crowds as continuous flows.
These approaches share similar advantages with SFM in terms of interpretability and computational efficiency, but they similarly rely on simplified assumptions that limit their ability to represent nuanced, context-dependent interactions.

Over the years, several extensions to the original SFM have attempted to address these limitations. For example, the Social Robot Force Model (SRFM)~\cite{agrawal2024evaluating} introduces an additional robot-induced repulsion force, while other studies have incorporated group cohesion forces~\cite{hossain2022sfmgnet}. The closest work to ours was done by Ratsamee et al.~\cite{ratsamee2015social}, which considers different pedestrian behaviors towards robots. However, these variants typically rely on manually tuning the parameters and training on datasets that lack explicit annotations of diverse pedestrian-robot interactions. Consequently, they fail to capture the nuances and context-dependent nature of pedestrian responses to robot navigation behavior. Similarly, other data-driven approaches such as recurrent neural network-based models~\cite{gupta2018social} and graph-based models~\cite{huang2019stgat}, although they improve prediction accuracy, suffer from high computational cost and lack of interpretability.

More recently, there has been a shift towards leveraging large language models (LLMs)~\cite{bae2024can} and diffusion-based methods~\cite{bae2024singulartrajectory} for trajectory prediction, enabling effective capture of stochasticity and diversity of human motion patterns. While these methods represent a significant advancement, they still rely heavily on datasets that lack fine-grained annotations of pedestrian responses specific to robot behavior.

\begin{figure*}[t!]
    \centering
    \begin{subfigure}[a]{0.32\textwidth}
         \centering
         \includegraphics[width=\textwidth]{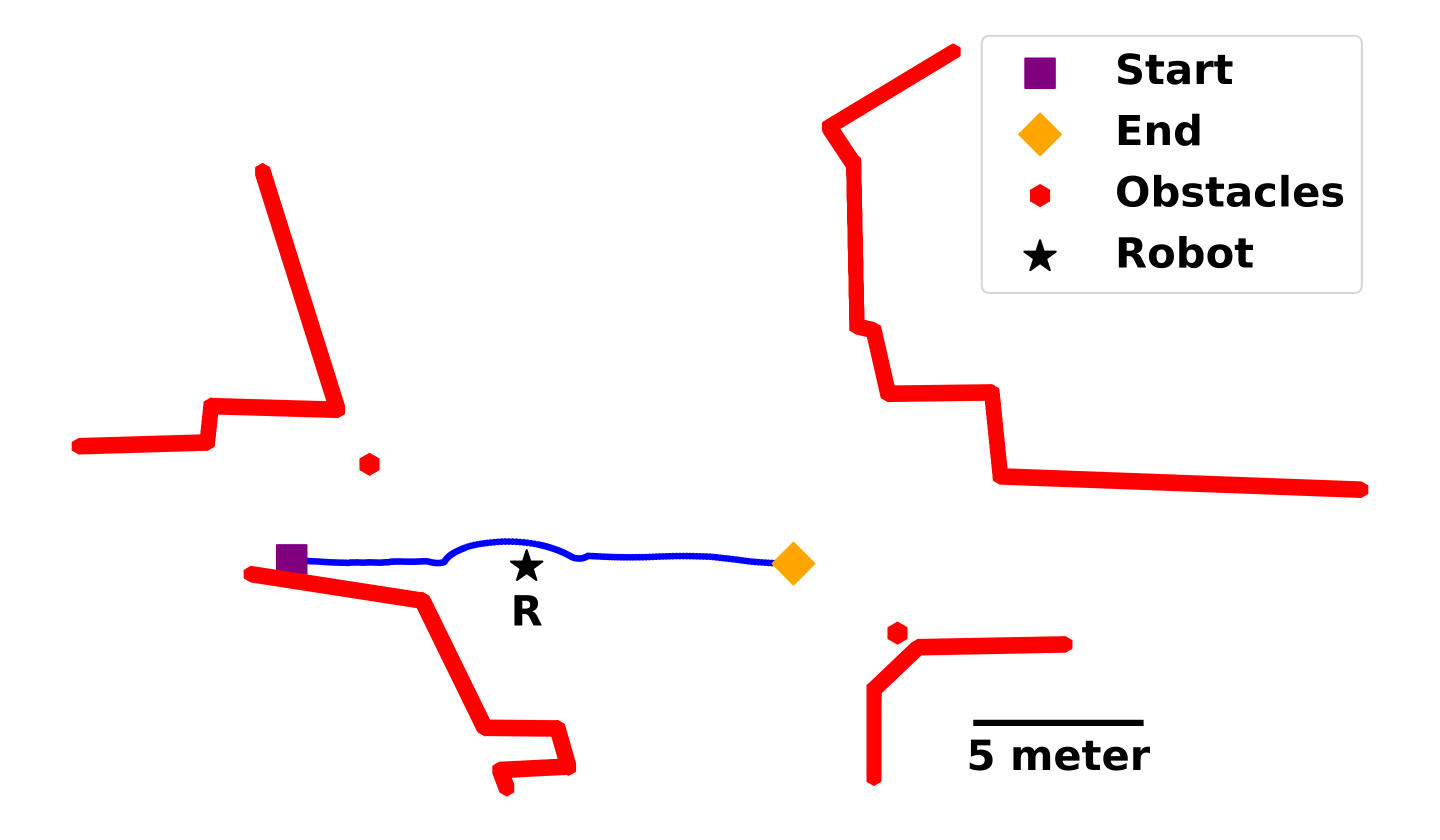}
         \caption{Avoidance behavior}
         \label{fig:avoid}
     \end{subfigure}
     \hfill
     \begin{subfigure}[a]{0.32\textwidth}
         \centering
         \includegraphics[width=\textwidth]{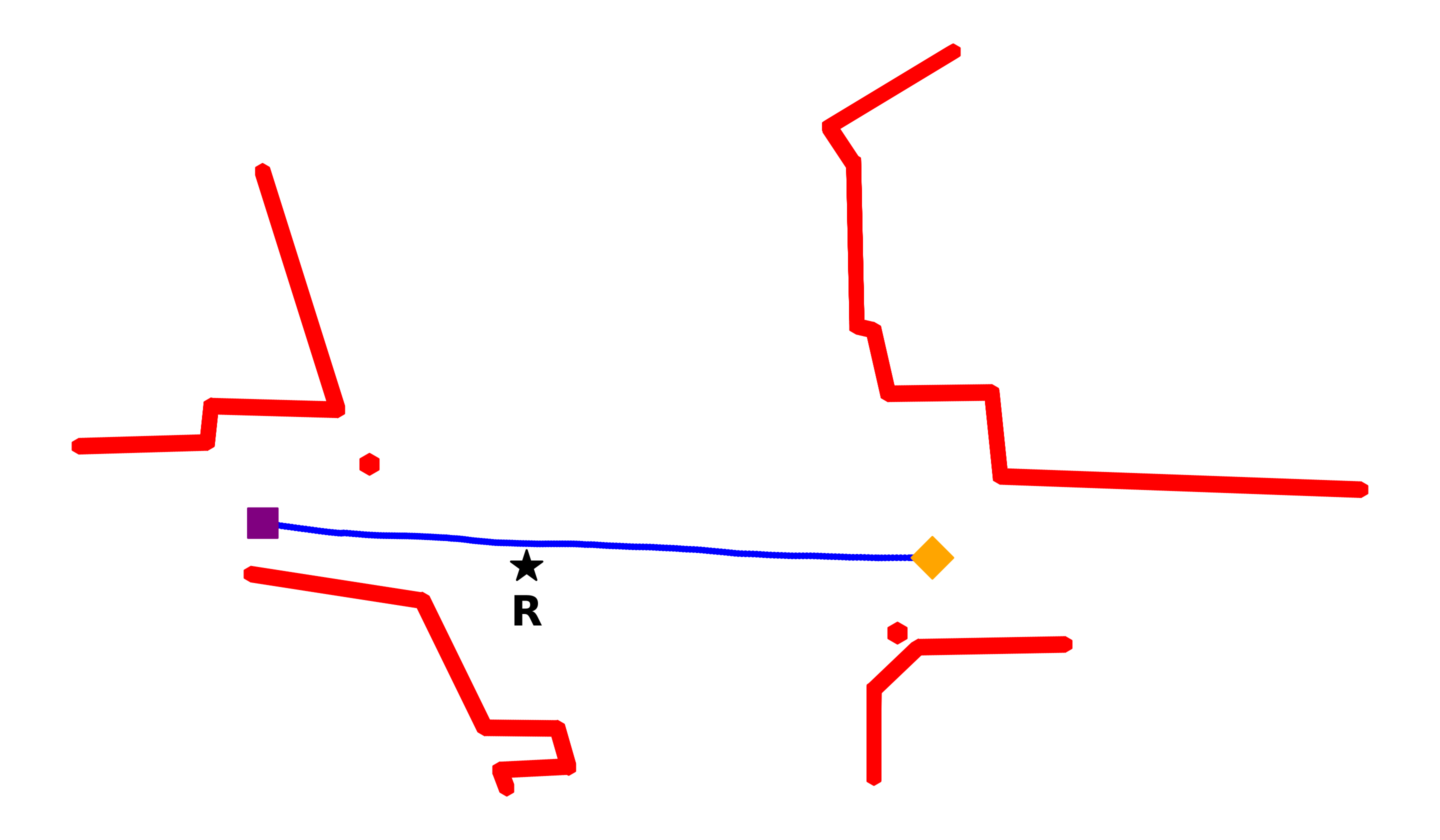}
         \caption{Neutral behavior}
         \label{fig:neutral}
     \end{subfigure}
     \hfill
     \begin{subfigure}[a]{0.32\textwidth}
         \centering
         \includegraphics[width=\textwidth]{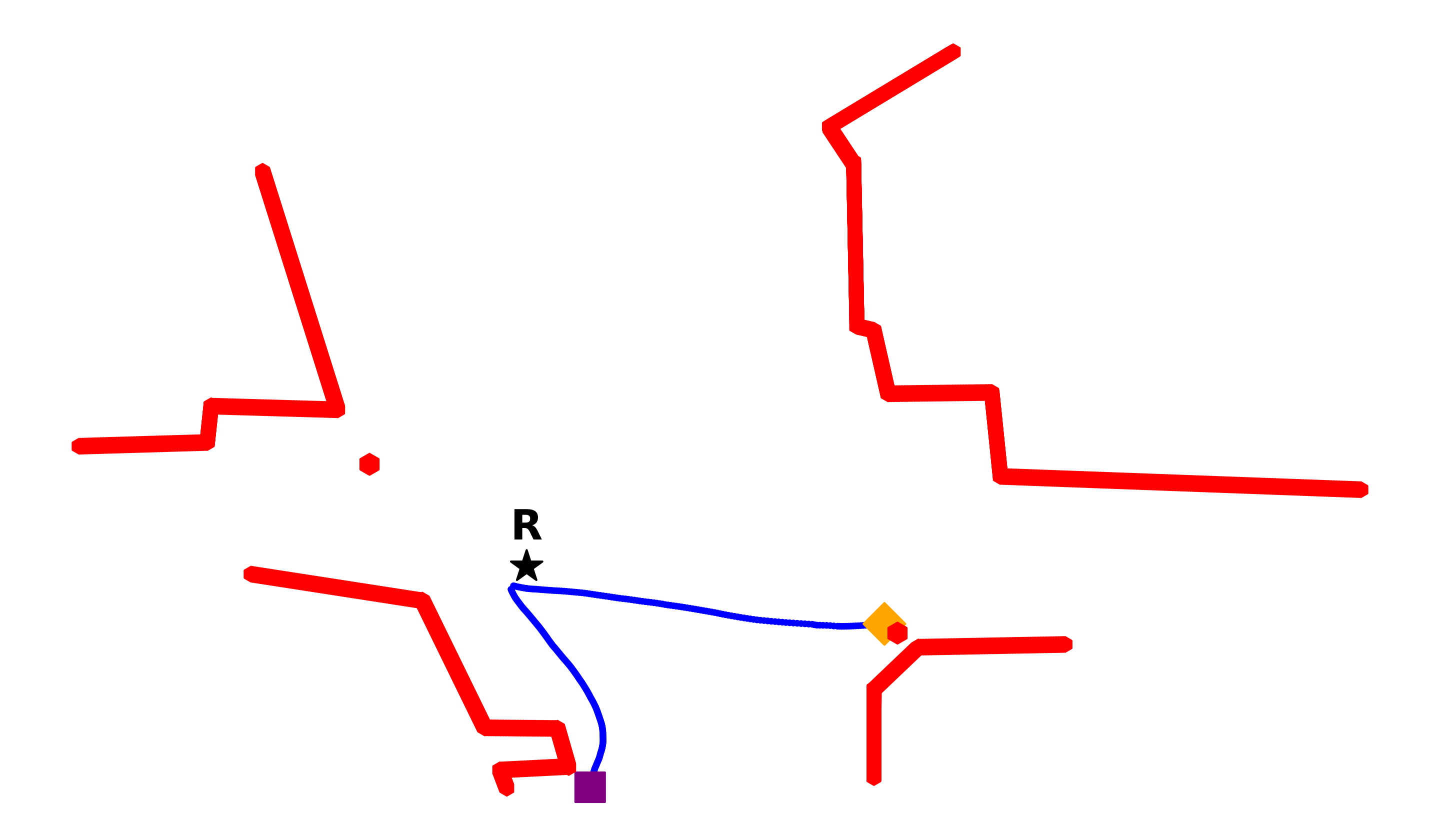}
         \caption{Attraction behavior}
         \label{fig:attract}
     \end{subfigure}
    \caption{
    Distinct pedestrian behaviors when close to robots taken from our PeRoI dataset: (a) The pedestrian clearly avoids the static robot (star) while walking toward their goal. (b) The pedestrian walks close to the robot without any noticeable change in trajectory direction. (c) The pedestrian deviates from their original path to approach the robot before resuming their goal-directed movement.}
    \label{fig:distinct_behavior}
    \vspace{-5px}
\end{figure*} 

\begin{table}[t]
\centering
\begin{tabular}{lp{0.75\linewidth}}
\toprule
\textbf{Category} & \textbf{Definition}\\
\midrule
\emph{Avoidance}  & Deviation from the nominal path to maintain distance from the robot (Fig~\ref{fig:avoid}). \\
\emph{Neutrality} & Negligible change in path or speed (Fig.~\ref{fig:neutral}). \\
\emph{Attraction} & Approach or orientation toward the robot (Fig.~\ref{fig:attract}). \\
\bottomrule
\end{tabular}
\caption{Behavioral categories of pedestrian responses to the robot.}
\label{tab:pedestrian_response_categories}
\end{table}

\begin{table*}[t]
\centering
\begin{tabular}{@{}p{0.14\textwidth}p{0.15\textwidth}p{0.20\textwidth}p{0.44\textwidth}@{}}
\toprule
\textbf{Field} & \textbf{Type} & \textbf{Units / Values} & \textbf{Description} \\
\midrule
Frame number & integer + timestamp & index, s & \textbf{-} Sequential frame index and wall-clock timestamp; dataset stored as a continuous sequence. \\
Pedestrian ID & integer & — & \textbf{-} Unique identifier per continuous track; re-entry into the scene receives a new ID. \\
Pedestrian $x$, $y$ position & float, float & m & \textbf{-} 2D position of pedestrian in scene coordinates relative to the origin. \\
Distance increment & float & m & \textbf{-} Displacement between consecutive frames; used to compute velocities. \\
Robot presence & boolean & \{0,1\} / \{false,true\} & \textbf{-} Indicates whether a robot is present in the scene. \\
Robot type & categorical & \{HSR, Go1, MPO700\} & \textbf{-} Robot platform when present; otherwise \texttt{N/A}. \\
Robot influence& categorical & \{attractive, neutral, avoidance\} & \textbf{-} Pedestrian response label when a robot is present. \\
Robot $x$, $y$ position& float, float & m & \textbf{-} 2D position of robot in scene coordinates relative to the origin. \\
\bottomrule
\end{tabular}
\caption{Dataset structure and field definitions.}
\label{tab:data_structure}
\end{table*}

\section{Robot-Pedestrian Influence Dataset}
\label{sec:PeRoI}
Our novel Pedestrian-Robot Interaction (PeRoI) dataset presented in this paper addresses the scarcity of large-scale, real-world pedestrian trajectory datasets that explicitly capture and annotate diverse responses to robots. 
It is based on the assumption that pedestrian behavior in a shared environment can be characterized by low-level motion features (position, heading, and velocity) and categorical behavior labels (avoidance, neutrality, attraction).
We explicitly designed the PeRoI dataset to enable the training of pedestrian models, which can predict diverse pedestrian responses to robots as shown in Sec.~\ref{sec:NeuRoSFM}. 

\begin{figure}[t]
    \centering
    \begin{subfigure}[b]{0.49\linewidth}
         \centering
         \includegraphics[width=\linewidth]{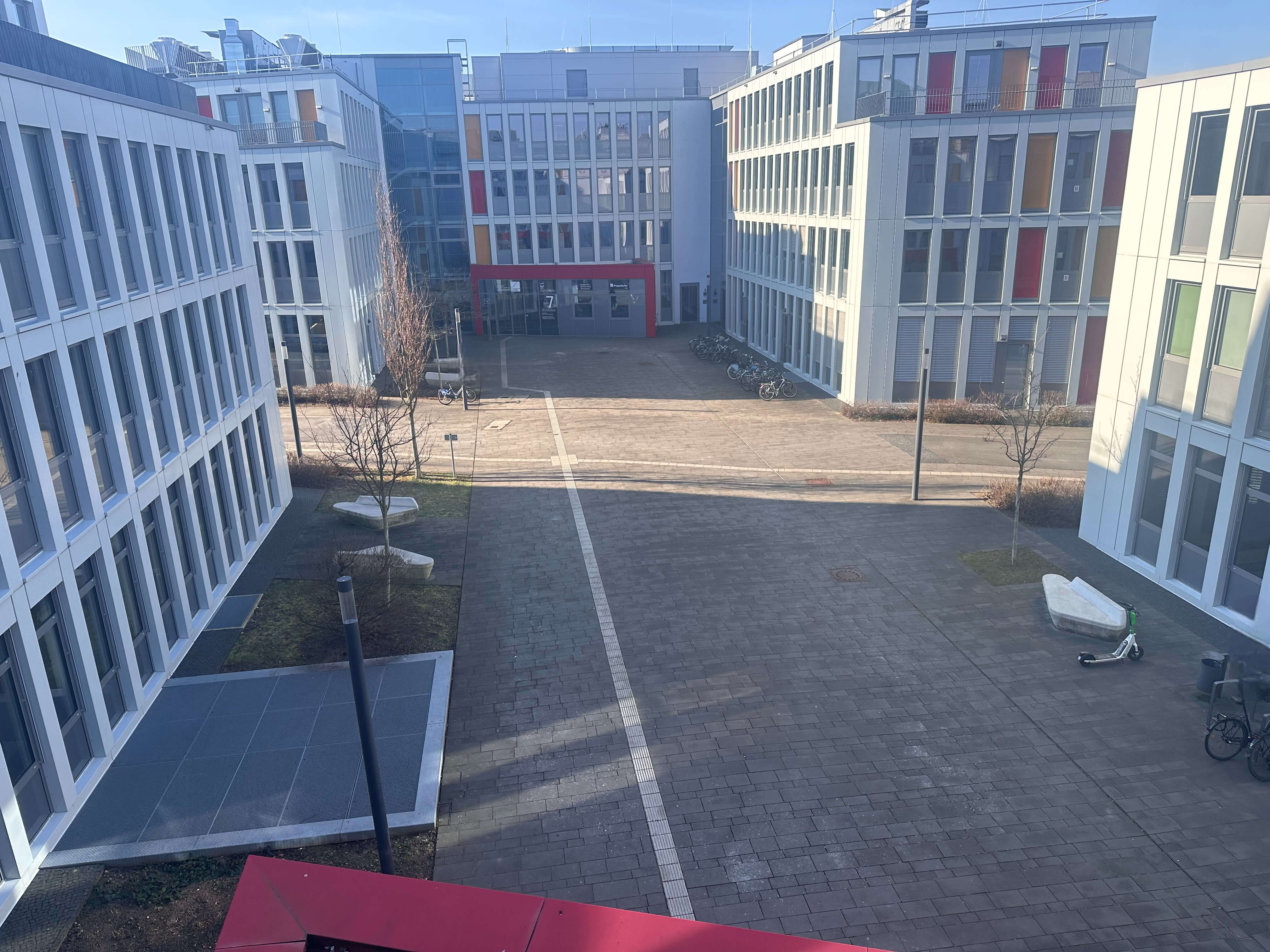}
         \caption{Environment 1}
         \label{fig:env1}
     \end{subfigure}
     \hfill
     \begin{subfigure}[b]{0.49\linewidth}
         \centering
         \includegraphics[width=\linewidth]{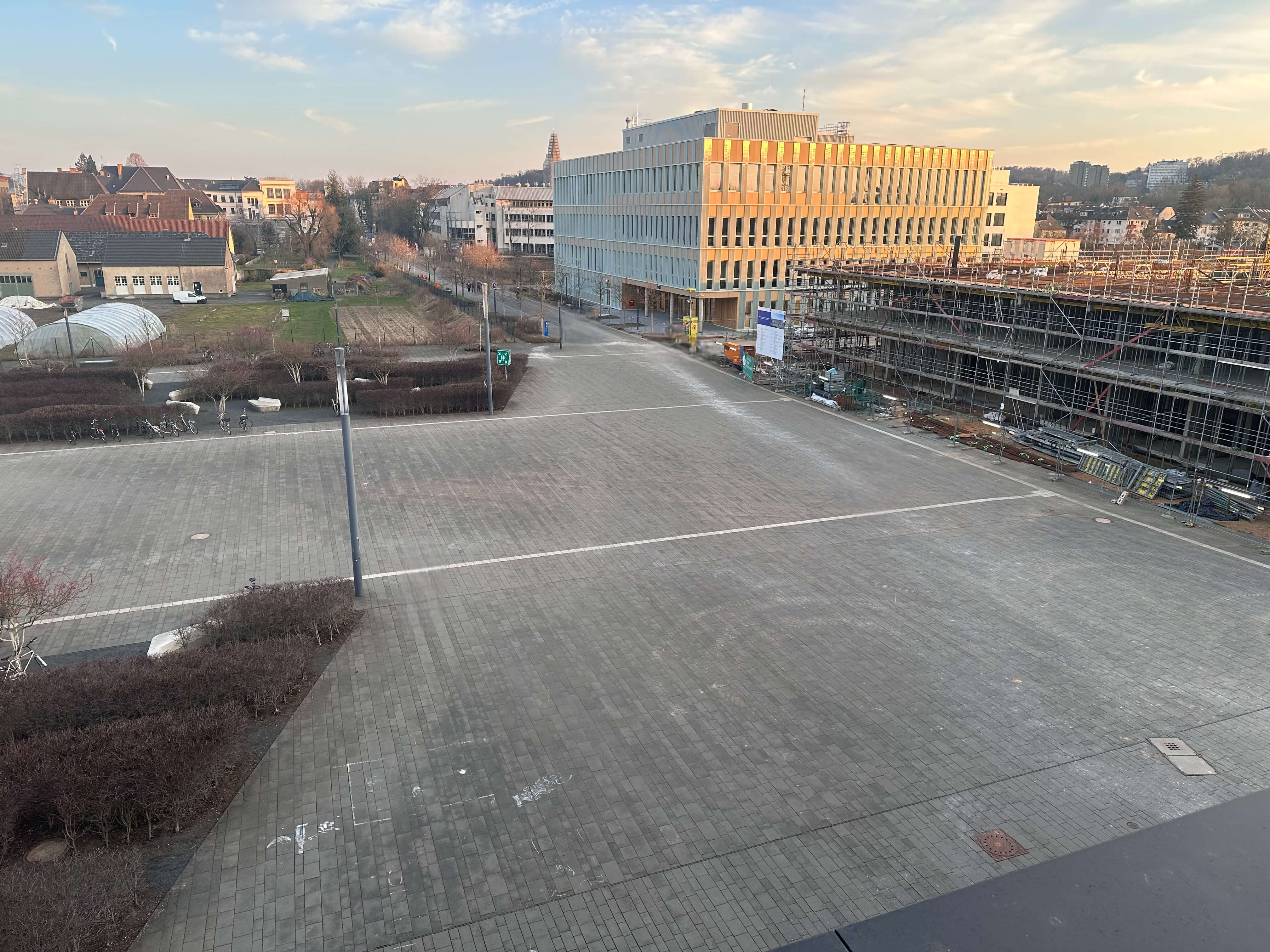}
         \caption{Environment 2}
         \label{fig:env2}
     \end{subfigure}
    \caption{Outdoor environments used for data collection. (a) A pathway crossing with two office building entries. (b) A larger university campus open space. }
    \vspace{-10px}
    \label{fig:outdoor_envs}
\end{figure}
\subsection{\textbf{Overview}}
We collected data in two outdoor settings (Fig.~\ref{fig:outdoor_envs}) with unstructured pedestrian flow:
\begin{itemize}
    \item \textbf{Environment 1} — a \(50\,\mathrm{m}\times 20\,\mathrm{m}\) pathway between two office buildings.
    \item \textbf{Environment 2} — a \(50\,\mathrm{m}\times 60\,\mathrm{m}\) open plaza on a university campus.
\end{itemize}

A fixed, overhead RGB camera from bird’s-eye perspective recorded the full scene in both environments at \(15\,\mathrm{Hz}\).
The elevated position minimized occlusions and preserved the global context, enabling the capture of long pedestrian and robot trajectories. 
Trajectories were recorded over a two-week period during peak working hours, capturing both individual and group motion. 
The resulting dataset comprises \(18{,}669\) trajectories spanning \(142\,\mathrm{h}\) across the two environments, of which \(16.45\%\) involve pedestrian–robot interactions. 
Pedestrian responses to the robot are annotated into three distinct behavioral categories, shown in Tab.~\ref{tab:pedestrian_response_categories}.
The full dataset follows the structure shown in Tab.~\ref{tab:data_structure} and will be published after acceptance. 
The dataset can be found on zenodo\footnote{https://doi.org/10.5281/zenodo.18876411}.


\subsection{\textbf{Data Collection Protocol}}
\label{subsec:data_collection}
To capture a comprehensive and diverse set of robot–pedestrian interactions, we deployed three distinct robot-presence conditions:
\begin{itemize}
    \item \textbf{PD (Pedestrians only):} Baseline pedestrian behaviour with no robot present. This baseline supports modeling of fundamental behaviors, that can be seen in traditional datasets, via forces such as goal attraction \(\vec{f}_a\), pedestrian–pedestrian repulsion \(\vec{f}_p\), obstacle repulsion \(\vec{f}_o\), and group dynamics \(\vec{f}_{\mathrm{gr}}\).
    \item \textbf{PD--SR (Pedestrians + Stationary Robot):} Pedestrian behavior with one of three different stationary robotic platforms placed at a fixed location, while pedestrian motion is recorded to assess reactions to a static robot.
    This provides data to model stationary robot-induced forces \(\vec{f}_{\mathrm{rs}}\), which vary with robot morphology and extend traditional datasets.
    \item \textbf{PD--MR (Pedestrians + Moving Robot):} 
    Since the Unitree Go1 received the most attraction in stationary case (Table~\ref{tab:pedestrian_responses}), we use it for further investigation in the mobile case. The robot is teleoperated along a predefined path within the observation area while pedestrian trajectories are tracked. This enables comparison between responses to stationary and moving robots and supports modeling of forces in the presence of moving robots~(\(\vec{f}_{\mathrm{rm}}\)).
\end{itemize}

\begin{figure}[t!]
    \centering
    \begin{subfigure}[b]{0.24\linewidth}
         \centering
         \includegraphics[width=\linewidth]{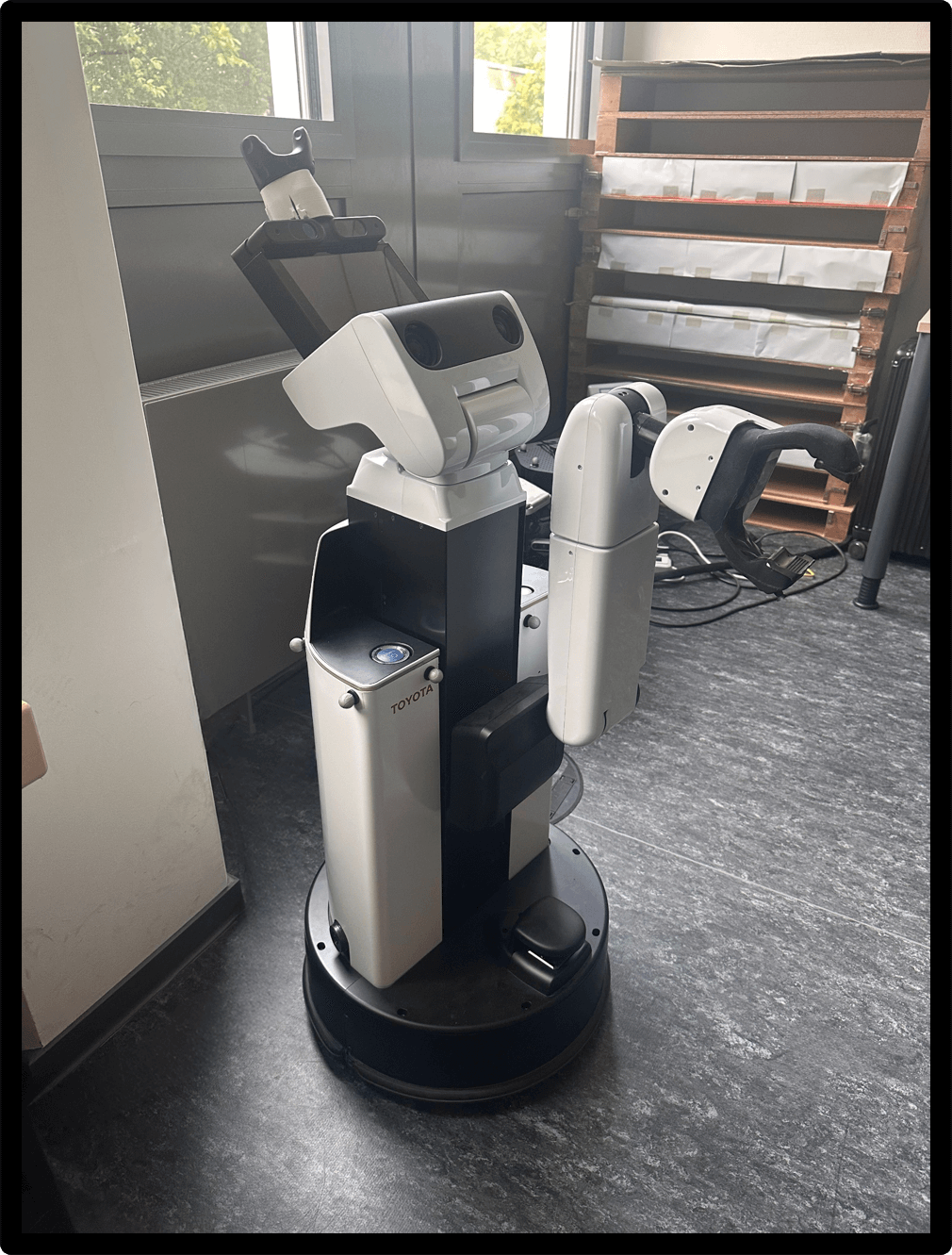}
         \caption{Toyota HSR}
         \label{fig:hsr}
     \end{subfigure}
     \hfill
     \begin{subfigure}[b]{0.49\linewidth}
         \centering
         \includegraphics[width=0.85\linewidth]{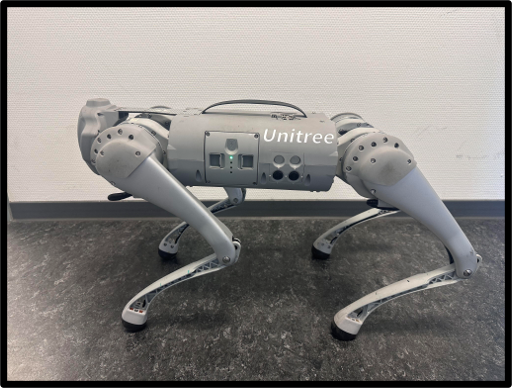}
         \caption{Unitree Go1}
         \label{fig:go1}
     \end{subfigure}
     \hfill
     \begin{subfigure}[b]{0.24\linewidth}
         \centering
         \includegraphics[width=\linewidth]{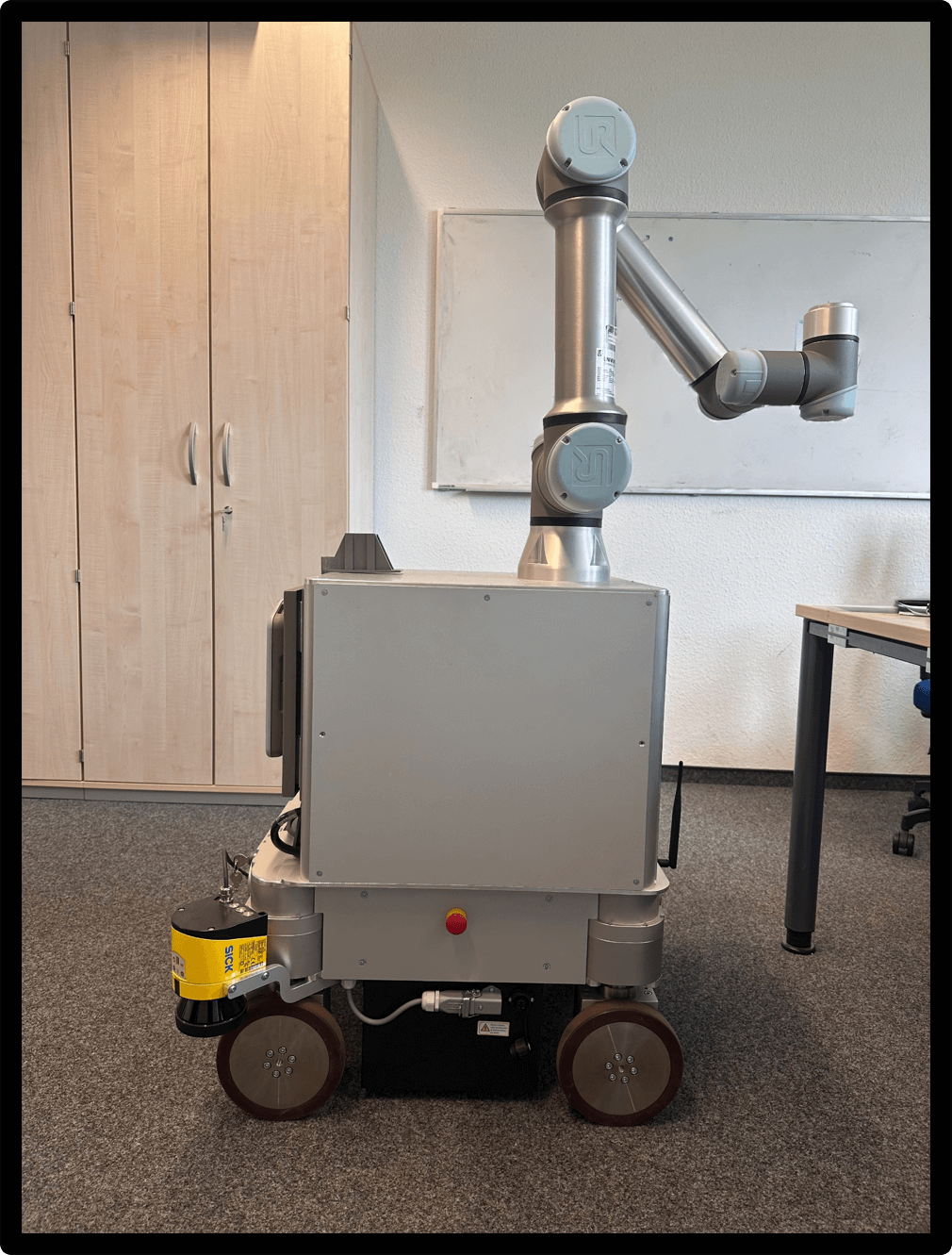}
         \caption{MPO700}
         \label{fig:steve}
     \end{subfigure}
    \caption{Robots used during the collection of the PeRoI dataset.}
    \vspace{-10px}
    \label{fig:robots}
\end{figure}

To allow analysis of morphology-dependent differences in human responses, we selected three robots representing distinct morphologies: a wheeled mobile manipulator used in social settings \mbox{(Toyota HSR, Fig.~\ref{fig:hsr})}, a quadruped \mbox{(Unitree Go1, Fig.~\ref{fig:go1})}, and an industrial mobile base \mbox{(Neobotix MPO700, Fig.~\ref{fig:steve})}.

Pedestrian detection and tracking were performed in real time using YOLOv11~\cite{yolo11_ultralytics}, with the tracked points mapped to 2D world coordinates via a planar homography. 
To protect privacy, no identifiable images or personal information were stored. 
Given the non-identifiable nature of the data, no special ethical approvals were required.
The data was collected through in-situ ethnographic observation of pedestrians in their natural surroundings to minimize observational bias.
Consequently, no formal recruitment or invitation process was used for the participants included in the dataset. 
The majority of observed pedestrians consisted of university students and staff, representing a younger demographic, while a small percentage included people from the older age groups. 
However, due to the unobtrusive nature of the data collection and associated privacy concerns, precise demographic information about the pedestrians was not obtained.

In total, the dataset contains \(\textbf{15{,}461}\) trajectories for PD, \(\textbf{2{,}948}\) for PD--SR (1,090 HSR; 837 MPO700; 1,021 Go1), and \(\textbf{260}\) for PD--MR, offering diverse interaction contexts for analysis.

\subsection{\textbf{Data Processing}}
A pedestrian trajectory is the spatiotemporal sequence of positions that describes an individual’s path. 
Formally, for a single pedestrian we define the trajectory
\(\mathcal{T}=\{\mathbf{x}_t\}_{t=1}^{T}\), with \(\mathbf{x}_t=(x_t,y_t)\in\mathbb{R}^2\) sampled at \(15\,\mathrm{Hz}\).
To ensure data quality and physical plausibility, trajectories were preprocessed according to the following criteria:

\begin{itemize}
    \item \textbf{Minimum-frame filter:} Trajectories shorter than \(10\) frames were removed, as false positives from the detector typically appear as brief and abrupt trajectories~\cite{host2020tracking}.
    \item \textbf{Spatial consistency:} Trajectories exhibiting unrealistic frame-to-frame displacements (i.e., “teleportation”) were discarded. 
    If the displacement was consistent with feasible walking velocities, e.g., due to detection failer, missing points were filled by linear interpolation.
    \item \textbf{Length threshold:} Trajectories with total arc length \(L<3.5\,\mathrm{m}\) were discarded, since very short movements provide limited information about interaction dynamics.
    \item \textbf{Speed constraints:} Trajectories with instantaneous speeds exceeding \(2.7\,\mathrm{m/s}\) were excluded to remove running, cycling, and other non-walking behaviours, focusing the dataset on pedestrian motion.
    \item \textbf{Behavioural consistency:} Looped or stationary patterns (e.g., lingering or circling) were removed, as they do not reflect goal-directed pedestrian movement relevant to interaction analysis.
\end{itemize}

For dataset labeling, it is essential to classify distinct human behaviors in response to the robot’s presence in the scene.
We perform the labeling of the trajectories manually where each trajectory is labeled based on the observed behavior of the pedestrian.

\section{Neural Robot Social Force Model (NeuRoSFM)}
\label{sec:NeuRoSFM}
Building upon the dataset, we propose a model that is capable of learning and predicting the distinct behavior of pedestrians in the presence of robots more accurately than the traditional Social Force Model (SFM)~\cite{helbing1995social}, that defines pedestrian motion as a result of cumulative attractive and repulsive forces acting on the pedestrian ($\vec{F}$), including a goal-directed attraction force ($\vec{f_a}$), repulsion from obstacles and walls ($\vec{f_o}$), repulsion from other pedestrians ($\vec{f_p}$), temporary attraction to objects and people in the environment~($\vec{f_t}$), and random fluctuations ($\epsilon_i$):

\begin{equation}
\label{eq:1}
\vec{F} = \vec{f_a} + \vec{f_o} +  \vec{f_p} +  \vec{f_t} + \epsilon_i
\end{equation}

\subsection{Extension of the Traditional Social Force Model}
Recently, Agrawal~\etal\cite{agrawal2024evaluating} demonstrated the need to augment the SFM with additional forces, such as a robot force, since the traditional SFM fails to capture the nuanced behaviors exhibited when pedestrians encounter robots. Additionally, we found that group forces significantly impact pedestrian behavior near robots, as they can influence an individual to move closer or farther away, independent of their intrinsic behavior. To address these complexities, we enhance the traditional SFM by learning a more complex robot force ($\vec{f_r}$) to model pedestrian-robot and a group force ($\vec{f_{gr}}$) to capture these social influences on human trajectories as described in the following.
\subsubsection{\textbf{Robot Force}}
The robot force is modeled as a repulsive force directed towards the pedestrian away from the robot. 
It follows a similar formulation as the repulsive force from other pedestrians as described by Helbing~\etal in the original SFM formulation~\cite{helbing1995social}. 
Mathematically, the robot force ($\vec{f_r}$) can be represented as:

\begin{equation}
\label{eq:robot_force}
\vec{f_r} =  - \nabla_{\vec{r}_{\alpha r}} \, V_{\alpha r}\!\left[b(\vec{r}_{\alpha r})\right]
\end{equation}

where, $- \nabla_{\vec{r}_{\alpha r}}$ is the gradient operator with respect to the relative position of the pedestrian ($\alpha$) and the robot ($r$), $V_{\alpha r}$ is the interaction potential between pedestrian and robot generally formulated as an exponentially decaying function of distance, and $b(\vec{r}_{\alpha r})$ is the Euclidean distance between the pedestrian and the robot. 
\subsubsection{\textbf{Group Force}}
The group force ($\vec{f_{gr}}$) follows the same formulation as shown in~\cite{hossain2022sfmgnet, ahmed2019investigating}. 
This force ensures that the pedestrian maintains a certain distance from the group centroid and does not move too far away from the group. The nature of the force~(attraction vs. repulsion) changes based on the distance of the pedestrian from the group centroid.
\subsubsection{\textbf{Combined Forces}}
We simplify our formulation of social forces by omitting the temporary attraction force towards other objects acting on the pedestrians due to the lack of significant observation of this behavior in our dataset. Therefore, Eq.~\ref{eq:1} finally extends to:


\begin{equation}
\label{eq:NeuRoSFM}
\vec{F} = \vec{f_a} + \vec{f_o} + \vec{f_p} + \vec{f_r} +  \vec{f_{gr}}  +\epsilon_i
\end{equation}

It is important to note that the robot force ($\vec{f_r}$) only models the repulsive behavior of pedestrians towards a robot. However, our observations from the PeRoI dataset reveal that repulsive behavior towards a robot is not the only response, as pedestrians may also show attraction or neutrality. To account for this, we model neutral behavior as a case in which the robot is treated similarly to an obstacle, i.e., without any additional individualized force being applied. In contrast, we model attractive behavior as a temporary change in the pedestrian's goal to the robot, after which the pedestrian resumes movement toward their original destination. Based on observations from the PeRoI dataset, where pedestrians exhibit this temporary goal-switching behavior in cases of attraction towards the robot, we set $n$ to the average attraction time of $\SI{5}{\second}$.



\subsection{Learning the Parameters of the NSFRM}

\begin{figure}[t]
    \centering
    \includegraphics[width=0.75\linewidth]{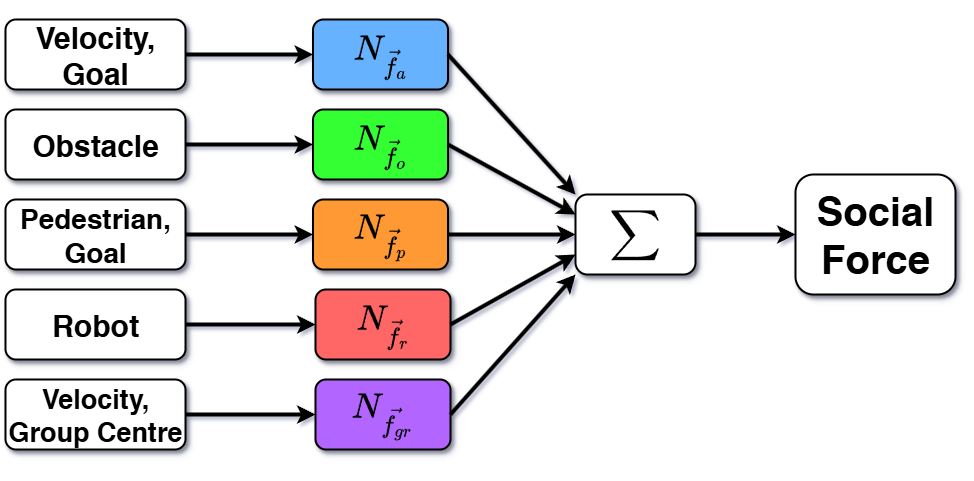}
    \caption{Structure of our proposed NeuRoSFM for pedestrian trajectory prediction. The input to NeuRoSFM include pedestrian velocity, goal direction, distance and direction to other pedestrians, distance and direction to the robot, and direction to the group centroid. These inputs are provided to five different networks, each of which outputs the force experienced due to the individual component. The final outputs are combined to get the resulting social force acting on the pedestrian.}
    \label{fig:NeuRoSFM_structure}
\end{figure}

The original SFM uses mathematical formulas to represent its individual forces, requiring extensive fine-tuning and strong expert knowledge for effective parameter tuning~\cite{helbing1995social}. Inspired by recent work on physics-infused and neural extensions of the SFM~\cite{zhang2022physics, hossain2022sfmgnet}, we replace these analytic formulations with neural network-based models. Specifically, we contribute a machine learning driven extension of the SFM that incorporates a robot-induced force component. This allows us to learn the individual forces accurately from the dataset without manually fine tuning the parameters.

Our proposed NeuRoSFM learns force components from the real-world data of the PeRoI dataset, thereby eliminating the need for manual parameter adjustment. To achieve this, we employ five separate neural networks, each dedicated to computing a specific type of force, as shown in Fig. \ref{fig:NeuRoSFM_structure}. The outputs of these individual networks are combined additively to yield the overall social force that drives pedestrian motion, similar to the traditional SFM. The individual networks in the NeuRoSFM are all multi-layer perceptrons (MLPs) and are as follows:
\subsubsection{\textbf{Goal Attraction Network, $N(\vec{f_a})$}} A twin-branched MLP that predicts goal-directed forces. One branch processes the pedestrian's velocity, while the other uses goal direction. Trained on straight-line trajectories, it ensures accurate goal-seeking behavior.
\subsubsection{\textbf{Obstacle Repulsion Network, $N(\vec{f_o})$}}
A two-stage MLP that takes the distance and unit direction vector to obstacles as input, and outputs repulsion forces 
Trained on pedestrian trajectories that demonstrate direct obstacle avoidance.
\subsubsection{\textbf{Pedestrian Repulsion Network, $N(\vec{f_p})$}}
Similar to $N(f_a)$ but incorporates inputs for pedestrian distance and direction. 
Due to the anisotropy of human perception and attention, it filters for individuals outside the pedestrian’s field of view and is trained on real and synthetic avoidance trajectories. The synthetic trajectories are generated using the SFM and help enrich the training data with cases that might be underrepresented in the dataset. This combination of real and synthetic data improves model robustness and enables better generalization across heterogenous pedestrian behaviors.
\subsubsection{\textbf{Robot Repulsion Network, $N(\vec{f_r})$}}
Similar in structure to the $N(f_o)$, this model predicts pedestrian repulsion from robots based on distance and direction. Trained on trajectories where pedestrians show evasive behavior near robots.
\subsubsection{\textbf{Group Cohesion Network, $N(\vec{f_{gr}})$}}
A twin-branched model that maintains pedestrian proximity to a group. 
One branch processes velocity in the direction of the goal, while the other uses the direction to the group centroid, outputting an attraction force towards the group.

\section{Experimental Evaluation}
\label{sec:evaluation}

The first goal of our evaluation is to demonstrate the quality and distinctiveness of the proposed PeRoI dataset. Thus, we first compare the various analytical results of our PeRoI dataset, followed by a comparison of our dataset with existing ones in the literature. Additionally, we evaluate the effectiveness of NeuRoSFM, showing that explicitly modelling diverse pedestrian responses to robots yields more accurate trajectory predictions. 

\begin{table*}[ht!]
\centering
\begin{tabular}{l c c c c c c c c c}
\toprule
\textbf{Dataset} & \textbf{Year} & \textbf{Pedestrian Trajectories} & \textbf{Robot Trajectories} &  \textbf{Avoidance Behavior} & \textbf{Neutral Behavior} & \textbf{Attraction Behavior} \\
\midrule
UCY          & 2007 & \cmark & \xmark & \cmark & \xmark & \xmark \\
ETH          & 2009 & \cmark & \xmark & \cmark & \xmark & \xmark \\
Stanford Drone & 2016 & \cmark & \xmark & \cmark & \xmark & \xmark \\
JRDB         & 2021 & \cmark & \cmark & \cmark & \xmark & \xmark \\
SCAND        & 2022 & \cmark & \cmark & \cmark & \xmark & \xmark \\
TBD          & 2024 & \cmark & \cmark & \cmark & \xmark & \xmark \\
PARSNiP      & 2025 & \cmark & \cmark & \cmark & \xmark & \xmark \\
\midrule
PeRoI (Ours)         & 2025 & \cmark & \cmark & \cmark & \cmark & \cmark \\
\bottomrule
\end{tabular}
\caption{Comparison of datasets for socially-aware robot navigation. Our PeRoI dataset contains distinct behavior labels of pedestrians towards robots} 
\label{tab:checkmark_comparison}
\end{table*}

\subsection{Robot-Pedestrian Influence Dataset}

\begin{table}[t]
    \centering
    \resizebox{\linewidth}{!}{
    \begin{tabular}{l|ccc}
        \toprule
        \textbf{Robot Type} & \textbf{Attraction (\%)} & \textbf{Avoidance (\%)} & \textbf{Avg. Distance (m)} \\
        \midrule
        HSR (Stationary) & 4.39 & 27.17 & 3.05 \\
        MPO700 (Stationary) & 1.6 & 33.95 & 3.26 \\
        Go1 (Stationary) & 7.82 & 26.39 & 3.24 \\
        Go1 (Moving) & 7.96 & 26.1 & 3.41 \\
        \hline
    \end{tabular}
    }
    \caption{Pedestrian responses to stationary and moving robots present in the PeRoI dataset. The responses are recorded in terms of attraction, avoidance, and average distance maintained. Attraction represents the percentage of pedestrians attracted to the specific robot. Avoidance represents the same feature but for pedestrians avoiding the robot. Average distance represents the distance maintained by pedestrians while navigating around the robot. Data shows that pedestrians were most attracted to the Unitree Go1 while avoiding the MPO700 the most.}
    \label{tab:pedestrian_responses}
    \vspace{-5px}
\end{table}

\subsubsection{\textbf{Interaction Responses in PeRoI}}
We compare interaction trajectories in PeRoI across stationary and moving robot conditions (PD--SR and PD--MR).
As shown in Tab.~\ref{tab:pedestrian_responses}, pedestrian responses depend strongly on the robot's embodiment: among pedestrians who interacted with the robot, the quadruped GO1 showed the highest attraction, whereas the industrial base MPO700 demonstrated the least. 
Moreover, the moving-robot condition yields the highest attraction rate overall, exceeding the corresponding stationary case. 
Based on the mean pedestrian--robot separation, the repulsive effect of a moving platform is similar in character but stronger than that of a stationary one, i.e., $\vec f_{rm} > \vec f_{rs}$.
We use these empirical results differentiate the robot--pedestrian repulsion~($\vec{f_r}$) in NeuRoSFM.

\begin{table}[t]
    \centering
    \begin{tabular}{l|ccc}
    \toprule
        \textbf{Dataset} & \textbf{Trajectories} & \textbf{RIT Trajectories} & \textbf{Percentage} \\ \midrule
        ETH & 750 & 0 & 0 \% \\
        JRDB & 1,786 & 28 & 1.57 \% \\
        PeRoI (Ours) & \textbf{18,669} & \textbf{3,071} & \textbf{16.45 \%} \\
    \hline
    \end{tabular}
    \caption{Comparison of common datasets in terms of number of trajectories with and without human-robot interaction. Our PeRoI dataset shows the highest percentage of trajectories where the human reacts to the presence of the robot in the scene compared to existing datasets such as the ETH~\cite{helbing1995social} and the JRDB~\cite{martin2021jrdb}. This provides more useful data for training models that account for pedestrian-robot interactions. }
    \vspace{-10px}
    \label{tab:dataset}
\end{table}

\begin{table*}[ht!]
\centering
\begin{tabular}{l | c c c c c c c}
\toprule
\textbf{Method \textbackslash ~Sets} & \textbf{ETH} & \textbf{HOTEL} & \textbf{UNIV} & \textbf{ZARA1} & \textbf{ZARA2} & \textbf{PeRoI} & \textbf{Average} \\
\midrule
DDL (training w/o PeRoI)                         & 0.26/0.50 & 0.15/0.35 & 0.29/0.58 & 0.16/0.29 & 0.13/0.22 & - & 0.20/0.39 \\
DDL (training w/ PeRoI)                        & \textbf{0.23/0.44} & \textbf{0.10/0.21} & \textbf{0.28/0.53} & 0.16/0.30 & \textbf{0.10/0.23} & \textbf{0.28/0.53} & \textbf{0.17/0.34} \\
\bottomrule
\end{tabular}
\caption{ADE/FDE with best-of-20 strategy for DDL~\cite{wang2024pedestrian} with and without the PeRoI dataset. ADE represents the average displacement error wrt. the ground truth in trajectory prediction across all the pedestrians. 
FDE represents the final displacement error, which is the error between the predicted goal position and the actual goal position. 
The first row represents evaluation of DDL without training on PeRoI dataset. The second row contains evaluation with PeRoI in the training set. Results show improved trajectory prediction performance of the DDL model with PeRoI in the training mix.}
\label{DDL_on_ETH_PeRoI}
\end{table*}

\subsubsection{\textbf{Literature Comparison}}
To compare PeRoI to prior work, Tab.~\ref{tab:checkmark_comparison} summarizes dataset-level features across the literature. 
As can be seen, PeRoI contributes a broader spectrum of human response annotations to robots during social navigation, spanning attraction/avoidance cues and embodiment-specific effects, with trajectories readily expressed in robot-centric frames.

\subsubsection{\textbf{Scale and Robot-Influenced Coverage}}
To evaluate the advantage of our PeRoI dataset in more detail, we compare it against the ETH~\cite{pellegrini2009you} and JRDB~\cite{martin2021jrdb} datasets in terms of total pedestrian trajectories, robot-influenced trajectories~(RITs), and the percentage of RITs in the whole dataset~(Table~\ref{tab:dataset}). 
ETH, a standard benchmark for pedestrian prediction, has 750 trajectories without human-robot interactions. 
JRDB contains 1{,}786 trajectories across indoor and outdoor scenes, of which 28 are RITs. 
In contrast, PeRoI offers 18{,}669 trajectories, including 3{,}071 RITs (16.45\%), exceeding both datasets in size and proportion of robot-influenced behavior. 
This higher RIT density supports more robust evaluation of models that incorporate robot forces ($\vec f_r$).

\subsubsection{\textbf{Velocity Distributions and Dynamics}}
\begin{figure}
    \centering
    \includegraphics[width=\linewidth]{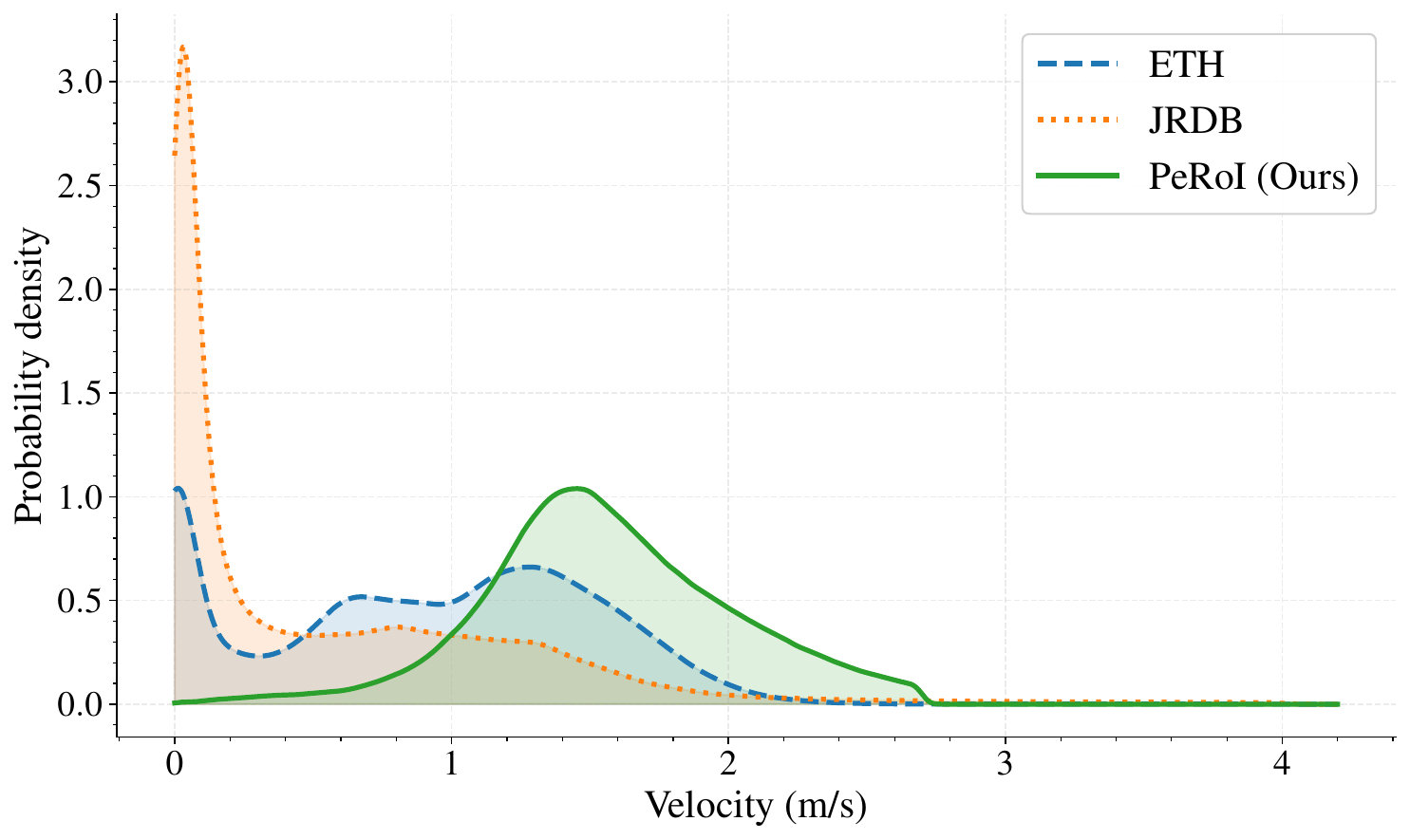}
    \caption{Velocity distribution of ETH, JRDB, and PeRoI datasets. Both the ETH and JRDB datasets exhibit a skewed velocity distribution, with a significant portion of pedestrians displaying null or very low velocities. In comparison, our PeRoI dataset  has a more uniform distribution of velocities amongst the tracked pedestrians. This is most likely due to the environment, where in case of ETH and JRDB there seem to exist more areas that cause pedestrians to stop moving while in case of PeRoI, the environment facilitates a more continuous motion. This broader and balanced distribution with a pronounced peak around natural walking speeds ($\sim1.2 m/s$) reflects more natural pedestrian-robot interaction contexts.}
    \label{fig:velocity_distribution}
\end{figure}

In addition, we analyze pedestrian speed distributions (\figref{fig:velocity_distribution}) to characterize the motion distributions across the datasets. 
In ETH, where no robot is present, the distribution is broad and multimodal. This means, that many trajectories cluster are near zero speed, e.g., due to people stopping or waiting, while a secondary peak appears around typical walking speeds~(about $1.3$–$1.5~\mathrm{m/s}$~\cite{chandra2013speed}). 
In JRDB, speeds are shifted further toward low values (median $0.25~\mathrm{m/s}$), which can be explained with confined layouts and the nearby robot encouraging caution and frequent halts. 
In contrast, PeRoI shows a narrow, unimodal distribution centred near $1.5~\mathrm{m/s}$; this suggests more consistent, sustained walking with fewer stops and less loitering, even in the robot’s presence. 
Further tests confirm that these differences are large and statistically significant according to the Mann–Whitney U Test.
In more detail, the test measures the degree of overlap between two distributions without assuming they are normally distributed, with a $\delta$ value of more than $0.474$ indicating a large effect. 
For our dataset we get $\delta=-0.61$ and $p<0.001$ against ETH and $\delta=-0.77$, $p<0.001$ compared to JRDB.
Together, these results indicate that PeRoI complements existing resources by capturing distinct robot-influenced dynamic behavior.

\subsubsection{\textbf{Trajectory-Prediction Benchmarking}}
To validate the dataset quality, we train a state-of-the-art dynamics-based deep learning model for pedestrian trajectory prediction~\cite{wang2024pedestrian}. 
Using ETH in the standard \mbox{\emph{best-of-20}} evaluation (train on four splits, test on the held-out split), we reproduce their first-row results in Table~\ref{DDL_on_ETH_PeRoI}. 
We omit robot influence in this case because ETH contains only natural pedestrian motion without robot interactions.
Incorporating PeRoI as an additional split and training on all-but-one split improves performance on all but one evaluation split (second row of Table~\ref{DDL_on_ETH_PeRoI}), where performance remains unchanged. 
Overall, adding PeRoI shows improved results, indicating that its data is at least on par with ETH for learning predictive dynamics. 

In summary, PeRoI offers a high proportion of robot-influenced trajectories among standard benchmarks, captures embodiment- and motion-dependent responses, and exhibits distinctive velocity profiles. 
Taken together, these properties make PeRoI an effective dataset for developing and evaluating navigation models that explicitly encode robot influences.

\subsection{Evaluation of the NeuRoSFM}

\begin{table}[t]
    \centering
    \renewcommand{\arraystretch}{1.2}
    \setlength{\tabcolsep}{8pt}
    \resizebox{\linewidth}{!}{
    \begin{tabular}{c|ccc}
    \toprule
        \textbf{Model \textbackslash ~Datasets} & \textbf{ETH} & \textbf{JRDB} & \textbf{PeRoI (Ours)} \\ \midrule
        NeuRoSFM (Ours) & \textbf{0.474} & \textbf{0.217} & \textbf{0.744} \\
        SRFM \cite{agrawal2024evaluating} &  0.616 & 0.336 & 1.117 \\ 
        SFM \cite{helbing1995social} & 0.616 & 0.412 & 1.118 \\ 
    \hline
    \end{tabular}
    }
    \caption{Comparison of different variations of the SFM in terms of \textbf{Average Displacement Error (ADE)} in meters. The results demonstrate that our NeuRoSFM achieves the lowest ADE across all datasets, highlighting the effectiveness of incorporating robot forces, group forces, and learning-based approaches for force prediction. Our dataset shows higher ADEs compared to others due to having longer trajectory lengths (139 frames in PeRoI compared to 18.1 in ETH and 58.67 in JRDB).}
    \vspace{-10px}
    \label{tab:ADE}
\end{table}

To validate the learned NeuRoSFM model, we compare its trajectory predictions against ground truth data from the ETH, JRDB, and PeRoI datasets. Prediction accuracy is quantified and reported in the form of average displacement error (ADE), defined as the average deviation of the predicted trajectory compared to the ground truth trajectory. Tab.~\ref{tab:ADE} compares the NeuRoSFM model alongside the optimization-based SRFM~\cite{agrawal2024evaluating} and the classical SFM \cite{helbing1995social}. The NeuRoSFM model is presented in three configurations as ablations: without robot force, without group force, and with both.

It is important to note that the JRDB dataset lacks group dynamics annotations, meaning that incorporating group force~$\vec{f_{gr}}$ has no effect. Similarly, the ETH dataset does not include robot presence, leading to identical results for SRFM and SFM, as the robot force $\vec{f{r}}$ is their only distinguishing factor.

The results show that NeuRoSFM consistently achieves the lowest ADE across all datasets, demonstrating its good predictive capability. Additionally, incorporating group force~($\vec{f_{gr}}$) further improves performance in datasets where group information is available, highlighting the importance of group dynamics. The use of robot force ($\vec{f_{r}}$) in SRFM also improves performance relative to SFM, while learning-based formulation of the forces in NeuRoSFM further enhances accuracy.

Overall, these findings confirm that integration of group forces, robot forces, and learning-based modeling significantly improves trajectory prediction when compared to traditional, manually tuned models.

\label{sec:conclusion}
\section{Conclusion}
We presented the PeRoI dataset and the NeuRoSFM model aimed towards advancing socially aware navigation. 
PeRoI offers large-scale, robot-conditioned pedestrian trajectories with response labels across different robot platforms and motion states.
The dataset captures pedestrian trajectories in three scenarios: no robot present, with a stationary robot, and with a moving robot, highlighting pedestrian avoidance, neutrality, and attraction behaviors throughout these cases. 
Unlike existing datasets, PeRoI explicitly annotates pedestrian responses to robots. 
Analyses show that moving robots elicit stronger repulsion than stationary ones and that responses vary by platform.
Fine-tuning a state-of-the-art predictor with the PeRoI dataset improves performance on most splits, indicating the data are informative rather than merely larger. 
Our novel prediction model NeuRoSFM learns robot- and group-induced forces and achieves accurate trajectory prediction across datasets, with ablations confirming the benefit of explicit robot terms. 
Our contributions enable the learning of models that encode robot influence and support more realistic evaluation of social navigation.
In the future, we plan to extend the dataset with more environments including both indoor and outdoor settings. We also plan to use additional sensors such as depth cameras and 3D lidar to collect a more comprehensive representation of the human-robot interactions in natural shared spaces.


\vspace*{-0.5em}
\printbibliography

\end{document}